\documentclass[twocolumn,showpacs,preprintnumbers,amsmath,amssymb]{revtex4}
\input psfig.sty

\usepackage{graphicx}
\usepackage{dcolumn}
\usepackage{bm}

\begin{document}

\title{A simplified approach for Chaplygin-type cosmologies}

\author{J. A. S. Lima} \email{limajas@astro.iag.usp.br} \author{J. V. Cunha}
\email{cunhajv@astro.iag.usp.br}

\affiliation{Instituto de Astronomia, Geof\'{\i}sica e Ci\^encias
Atmosf\'ericas, USP, 05508-900 S\~ao Paulo, SP, Brasil}

\author{J. S. Alcaniz}
\email{alcaniz@on.br} \affiliation{Departamento de Astronomia,
Observat\'orio Nacional, 20921-400 Rio de Janeiro, RJ, Brasil}

\date{\today}

\begin{abstract}

A new class of accelerating cosmological models driven by a
one-parameter version of the general Chaplygin-type equation of
state is proposed. The simplified version is naturally obtained
from causality considerations with basis on the adiabatic sound
speed $v_S$ plus the observed accelerating stage of the universe.
We show that very stringent constraints on the unique free
parameter $\alpha$ describing the simplified Chaplygin model can
be obtained from a joint analysis involving the latest SNe type Ia
data and the recent Sloan Digital Sky Survey measurement of baryon
acoustic oscillations (BAO). In our analysis we have considered
separately the SNe type Ia gold sample measured by Riess et al.
(2004) and the Supernova Legacy Survey (SNLS) from Astier et al.
(2006). At 95.4\% (c.l.), we find for BAO + \emph{gold} sample,
$0.91 \leq \alpha \leq 1.0$ and $\Omega_{\rm{M}}=
0.28^{+0.043}_{-0.048}$ while BAO + SNLS analysis provides $0.94
\leq \alpha \leq 1.0$ and $\Omega_{\rm M}=0.27^{+0.048}_{-0.045}$.

\end{abstract}

\maketitle

\section{Introduction}

The impressive convergence of recent observational facts along
with some apparently successful theoretical predictions seem to
indicate that the simple approach provided by the standard cold
dark matter (CDM) model is insufficient to describe the present stage of
our universe. From these results, the most plausible picture for
our world seems to be a nearly flat scenario dominated basically
by CDM and an exotic component endowed with large negative
pressure, usually named dark energy. Despite the good
observational indications for the existence of these two
components, their physical properties constitute a completely open
question at present, which gives rise to the so-called dark matter
and dark energy problems (see \cite{rev123} for a recent review on this topic).

Among the many candidates for the dark energy component, a very
interesting one was suggested by Kamenshchik {\it et al.}
\cite{kamen} and developed by Bili\'c {\it et al.} \cite{bilic}
and Bento {\it et al.} \cite{bento}. Such an exotic fluid, named
generalized Chaplygin gas (C-gas), can be macroscopically
characterized  by the equation of state (EoS)
\begin{equation}\label{eq1}
p_C = -A/\rho_C^{\alpha},
\end{equation}
where $\alpha$ = 1 and $A$ is a positive constant related to the
present-day Chaplygin adiabatic sound speed, $v^2_s = \alpha
A/\rho_{C_{o}}^{1 + \alpha}$ ($\rho_{C_{o}}$ stands for the
current C-gas density). In actual fact, the above equation for
$\alpha \neq 1$ constitutes a generalization of the original C-gas
EoS proposed by Bento {\it et al.} in Ref. \cite{bento}.

In the last few years, the possibility of describing the unknown
dark energy component using the C-gas-type EoS above has provoked
a considerable debate in the literature. Theoretical connections
between the C-gas and string theory, supersymmetric
generalizations \cite{hope,jackiw}, self-interacting \cite{zhu}, and even a tachyonic fluid
representation \cite{Bena02} has also been investigated.  Another
interesting feature of the above EoS comes from the fact that the
C-gas becomes pressureless at high redshifts, which suggests a
possible unification scheme for the cosmological ``dark sector",
an interesting idea which has  been considered in different
contexts \cite{quartessence}.

Observational aspects of the above C-gas scenarios have also been
largely investigated in the literature. Cosmological tests
involving type Ia supernovae (SNe Ia) data \cite{fabris,alcaniz}, the
shape of the matter power spectrum \cite{avelino}, statistical
properties of gravitational lenses \cite{dev}, the age of the
Universe \cite{jailson},  cosmic microwave background (CMB)
measurements \cite{bento1,bert1,fin,finelli}, galaxy clusters
X-ray \cite{CAL04}, and gamma-ray bursts data \cite{bertgr}  have been discussed. In general, to
perform such analyses, besides the present value of the C-gas
density parameter ($\Omega_C$), the above barotropic EoS implies
that one needs to constrain two additional free parameters,
namely, $A$ and $\alpha$. Therefore, in the context of the
Friedman-Robertson-Walker (FRW) cosmologies with CDM
plus a C-gas, there are at least 4 parameters to be constrained by
the data. Actually, this number can be reduced to 3 if one assumes
a flat geometry, i.e., $\Omega_{\rm{M}}  = 1 - \Omega_C$ or if a
unified dark matter/energy picture involving only the C-gas and
baryons is assumed from the very beginning (in this case, the
baryonic density ($\Omega_b$) may be fixed a priori by using, for
instance, nucleosynthesis \cite{Steigmann} or the recent Cosmic Microwave
Background (CMB) observations \cite{wmap}). However, even in
this latter cases, there are so many parameters to be constrained
by the data, that a high degree of degeneracy on the parametric
space becomes inevitable.

Many generalizations of the  original C-gas
\cite{ZhWuZh06,ChLa05,MaHa05,Wang05,GuZh05}, or even of its
extended version \cite{SeSc05} have appeared in literature. In
these cases, the number of free parameters is usually increased,
and, as consequence, the models become mathematically richer
although much less predictive from a physical viewpoint. In this
work by following the opposite direction, we propose a simplified
version for the generalized C-gas-type EoS which diminishes one of
its free parameters. By an additional physical condition, the
allowed range of the remaining parameter is also restricted a
priori, which makes not only the relevant parametric space
bi-dimensional but also (and more important) the model more easily
discarded or confirmed by the present set of observations since
the range of its free parameter is physically limited from
causality considerations. We test the viability of this simplified
C-gas approach by discussing the constraints imposed from current
SNe Ia observations and Large Scale Structure (LSS) data.

\section{A simplified C-gas scenario}

Let us consider a homogeneous and isotropic Universe whose energy
components are cold dark matter plus the generalized C-gas fluid.
Since both components are separately conserved, by inserting Eq.
(1) into the energy conservation law $\dot{\rho}_{C} = -3H
(\rho_{C} + p_C$), one obtains the following expression for the
density of the C-gas \cite{bento,quartessence,CAL04}
\begin{equation}\label{eq3}
\rho_{C} = \rho_{C_{o}}\left[A_s + (1 - A_s)a^{3(1 +
\alpha)}\right]^{\frac{1}{1 + \alpha}},
\end{equation}
where $a(t)$ is the cosmological scale factor and $A_s =
A/\rho_{C_{o}}^{1 + \alpha}$ is a convenient dimensionless
constant (as usual, the subscript ``0" denotes present-day
quantities). As one may check, the above C-gas evolving in the FRW
metric can be modeled as a \emph{quintessence}, that is, a scalar
field model described by an ordinary Lagrangian density, ${\cal
L}_{\phi}= {\frac{1}{2}}{\dot\phi}^{2} - V(\phi)$, with the
following potential
\begin{eqnarray}\label{eq3a}
V(\phi) &=&\frac{1}{2}\rho_{C_o} {A_s^{\frac{1}{\alpha +
1}}}\{[\cosh \sqrt{6\pi}m_{pl}(\alpha + 1)\phi)]^{\frac{2}{\alpha
+ 1}} \nonumber
\\&& + \,\, [\cosh\sqrt{6\pi}m_{pl}(\alpha +
1)\phi]^{-\frac{2\alpha}{\alpha + 1}}\},
\end{eqnarray}
where $m_{pl}$ is the Planck mass.

In a flat geometry, the Friedmann equation for a conserved C-gas
plus cold dark matter is given by \cite{CAL04}
\begin{eqnarray}\label{eq4}
{\cal{H}} & = & {\Omega_{\rm{M}}{a}^{-3} + 
\Omega_{C}[A_s + (1 - A_s) a^{-3(\alpha + 1)}]^{\frac{1}{\alpha +
1}}},
\end{eqnarray}
where ${\cal{H}} = H(a)^2/H_0^2$. Note that besides the Hubble
parameter $H_0$ we still have 3 additional parameters in this case
($\alpha, A_s, \Omega_{\rm{M}}$), even using the flat condition
$\Omega_C  = 1 - \Omega_{\rm{M}}$. Therefore, an interesting
question to be answered at this point is how to reduce the C-gas
parameters based on reasonable physical constraints?

\begin{figure*}[t]
\centerline{\psfig{figure=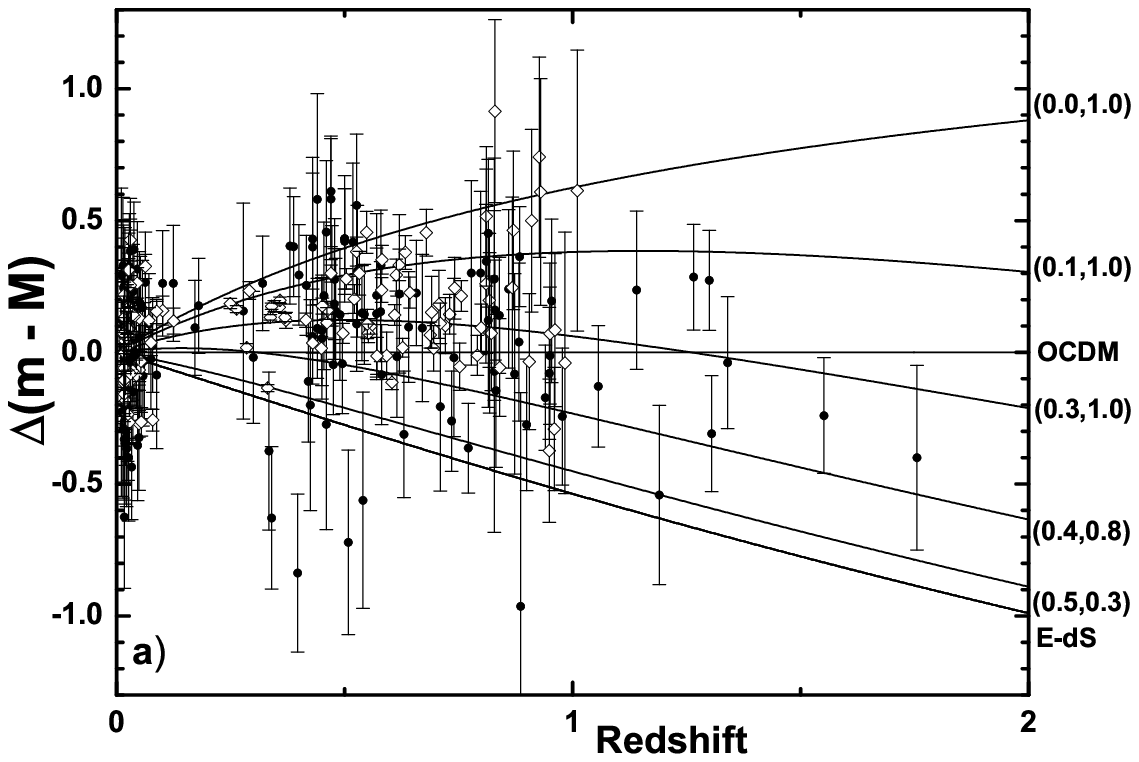,width=2.6truein,height=2.9truein} 
\psfig{figure=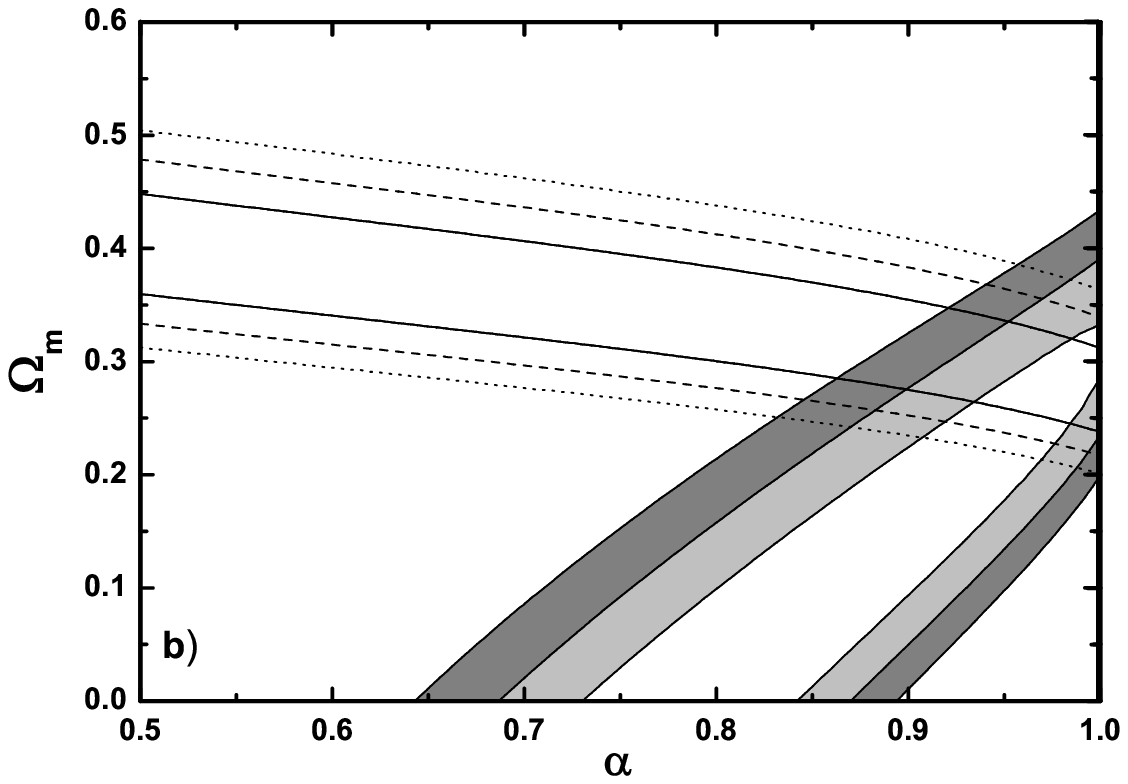,width=2.5truein,height=3.1truein}
\psfig{figure=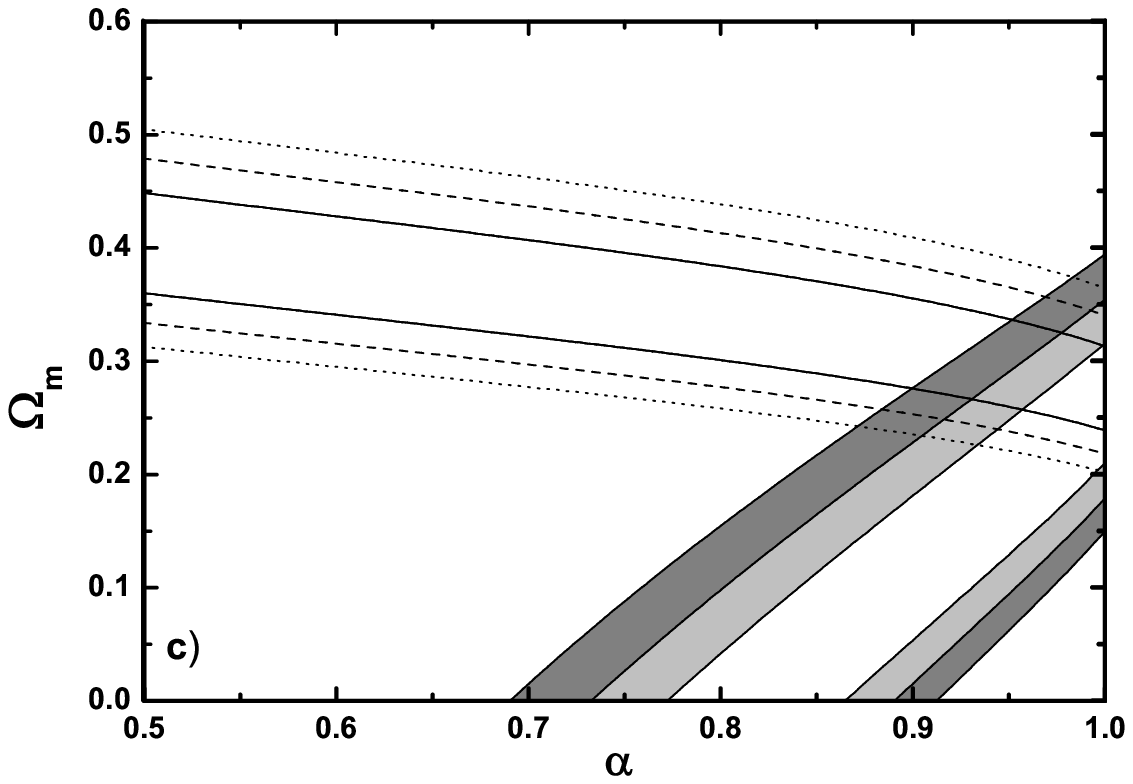,width=2.5truein,height=3.1truein}
\hskip 0.1in} 
\caption{Supernova results. Panel (a) displays the residual magnitude with respect to an empty ($\Omega_{\rm{T}}=0$) universe for the HZST (filled circles) and SNLS (open circles) samples, respectively. The solid curves are the predictions for the SG-gas-type models characterized by ($\Omega_{\rm M}, \alpha$). For comparison, we also display the predictions of the open cold dark matter (OCDM) and Einstein-de Sitter Universes. Panels (b) and (c) show 68.3\%, 95.5\% and 99.73\% confidence contours on the space parameter ($\Omega_{\rm M}, \alpha$) from the HZST and SNLS supernova data, respectively. The horizontal dotted and dashed lines correspond to the constraints arising from the SDSS baryon acoustic oscillations detection. For SNe Ia data alone, if $\alpha$ is greater than 0.91, values of $\Omega_{\rm M}$ smaller than 0.15 are ruled out by the two samples. Constraints from BAO contribute to increase the allowed values of $\Omega_{\rm M}$.} 
\end{figure*}

In order to answer the above question, we first notice that the
dimensionless constant $A_s$ appearing in the above expressions
encodes the basic information coming from the original parameter
$A$ [see Eq. (1)]. On the other hand, the Chaplygin adiabatic sound speed reads
\begin{equation}\label{eq5}
v_s^{2} = \frac{dp}{d\rho}= \alpha A/\rho_{C}^{1 + \alpha},
\end{equation}
which must positive definite for a well-behaved gas (zero in the
limit case of dust). Note also that the present day Chaplygin
adiabatic sound speed is $v_{so}^{2} = \alpha A/\rho_{C_{o}^{1 +
\alpha}}$ or, equivalently, 
\begin{equation}\label{eq6}
v_{so}^{2} = \alpha A/\rho_{C_o}^{1 + \alpha} = \alpha A_s.
\end{equation}
Therefore, if the $A_s$ parameter is a function of $\alpha$, the
number of free parameters is naturally reduced, and, as an extra
bonus, the positiviness of $v_s^{2}$ at any time, as well as its
thermodynamic stability, is naturally guaranteed. Among many possible relations (e.g., $A_s = \alpha^n$), clearly the simplest choice is $A_s = \alpha$ ($n = 1$). In this case, $v_{so}^{2} =
\alpha^{2}$, or more generally, $v_s^{2} =
{\alpha}^{2}(\rho_{Co}/\rho)^{\alpha}$. Note also that, since the
light speed is a natural cutoff for the sound propagation, it
follows that $v_{so}=|\alpha|\leq 1$, thereby restricting $\alpha$
to the interval [-1,1]. An additional constraint can still be
imposed to this parameter. In fact, with $A_s=\alpha$, the
simplified C-gas EoS (1) becomes
\begin{equation}\label{eq7}
p_C = -\alpha
\rho_{Co}\left(\frac{\rho_{Co}}{\rho_{C}}\right)^{\alpha},
\end{equation}
so that a negative pressure is obtained only for positive values
of $\alpha$. In other words, this accounts to saying that the
combined requirements from causality along with the observed
accelerating stage of the Universe limit naturally the  parameter
$\alpha$ to the interval $0 < \alpha \leq 1$.

Note that the the  simplified Chaplygin gas above (from now on
SC-gas) preserves the unifying character of the original C-gas,
i.e., it behaves as a pressureless fluid (nonrelativistic matter)
at high-$z$ while, at late times,  it approaches the quintessence
behavior, which now is fully characterized by the $\alpha$
parameter (for a unified dark matter/dark energy description of the above scenario, see \cite{newCG}). However, note also that, even in this limiting case,
the sound speed is positive. In other words, the Universe
evolution resembles the one driven by a quintessence component but
the thermodynamic behavior does not present the pathologies of
such scenarios.

In this simplified approach, Eq.(\ref{eq4}) is rewritten as
\begin{eqnarray}\label{eq8}
{\cal{H}} & = & {\Omega_{\rm{M}}{a}^{-3} + 
\Omega_{C}[\alpha + (1 - \alpha) a^{-3(\alpha +
1)}]^{\frac{1}{\alpha + 1}}},
\end{eqnarray}
so that the parameter $\alpha$ is actually the unique unknown
constant related to this SC-gas model. In what follows, we
confront this simplified approach with the most recent SNe Ia and
Large Scale Structure (LSS) data.

\section{Observational Constraints}

\subsection{SNe Ia}

Let us first investigate the bounds arising from SNe Ia
observations on the SC-gas scnario described above. To this end we
use the most recent SNe Ia observations, namely, the High-Z SN
Search (HZS) Team \cite{Riess04} and the Supernova Legacy Survey
(SNLS) Collaboration data \cite{snls}.

The so-called \emph{gold} sample from the HZS team is a selection
of 157 SNe Ia events distributed over the redshift interval $0.01
\lesssim z \lesssim 1.7$, and constitutes the compilation of the
best observations made so far by them and by the Supernova
Cosmology Project plus 16 new events observed by Hubble Space
Telescope (HST). The current data from SNLS collaboration
correspond to the first year results of its planned five year
survey. The total sample includes 71 high-$z$ SNe Ia in the
redshift range $0.2 \lesssim z \lesssim 1$ plus 44 low-$z$ SNe Ia.
This data set is arguably (due to multi-band, rolling search
technique and careful calibration) the best high-$z$ SNe Ia
compilation to date, as indicated by the very tight scatter around
the best fit in the Hubble diagram and a careful estimate of
systematic uncertainties. Another important aspect to be
emphasized on the SNLS data is that they seem to be in a better
agreement with WMAP results than the \emph{gold} sample (see,
e.g., \cite{paddy} for a discussion). The two SNe Ia samples are
illustrated on a residual Hubble Diagram with respect to the empty
universe model ($\Omega_{\rm{T}} = 0$) in Fig. 1a.

The predicted distance modulus for a supernova at redshift $z$,
given a set of parameters $\mathbf{p}$, is
\begin{equation} \label{dm}
\mu_p(z|\mathbf{p}) = m - M = 5\mbox{log} d_L + 25,
\end{equation}
where $m$ and $M$ are, respectively, the apparent and absolute
magnitudes, the complete set of parameters is $\mathbf{p} \equiv
(H_o, \Omega_{\rm{M}}, \alpha)$ and $d_L$ stands for the
luminosity distance (in units of megaparsecs),
\begin{equation}
d_L = c(1 + z)\int_{x'}^{1} {dx \over
x^{2}{\cal{H}}(x;\mathbf{p})},
\end{equation}
with $x' = (1 + z)^{-1}$ being a convenient integration variable
and ${\cal{H}}(x; \mathbf{p})$ the expression given by Eq.
(\ref{eq8}).

\begin{figure*}[t]
\centerline{\psfig{figure=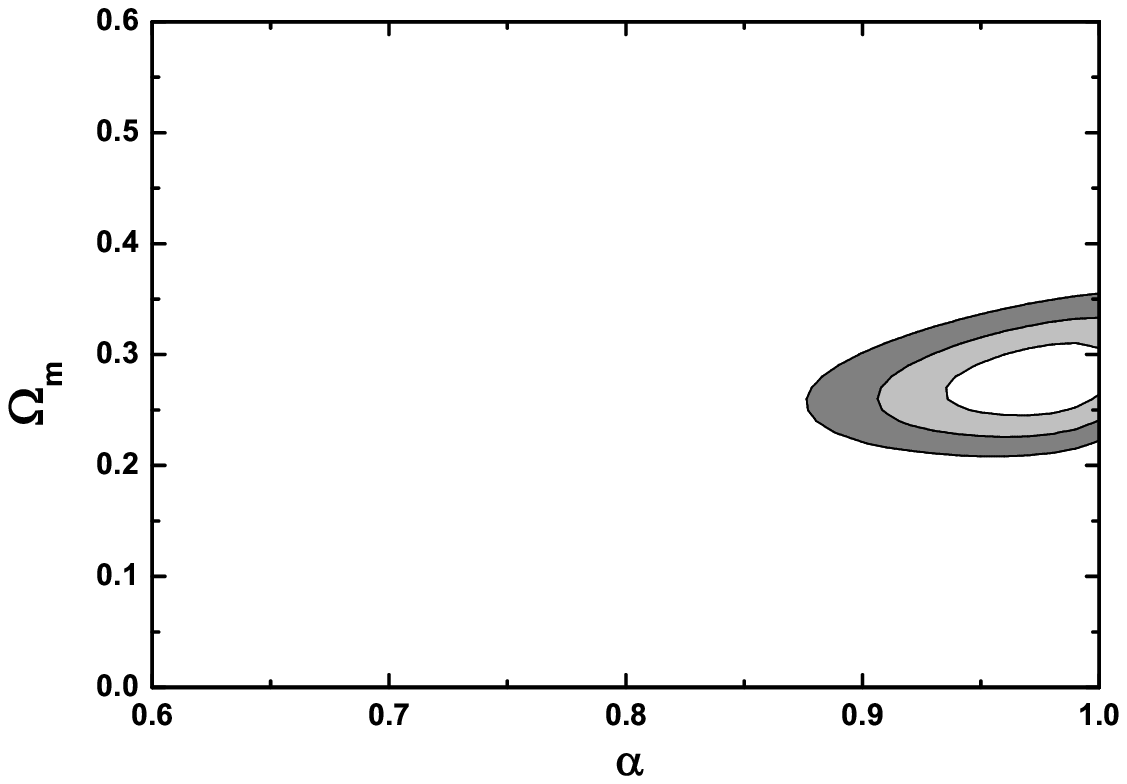,width=3.4truein,height=2.8truein}
\psfig{figure=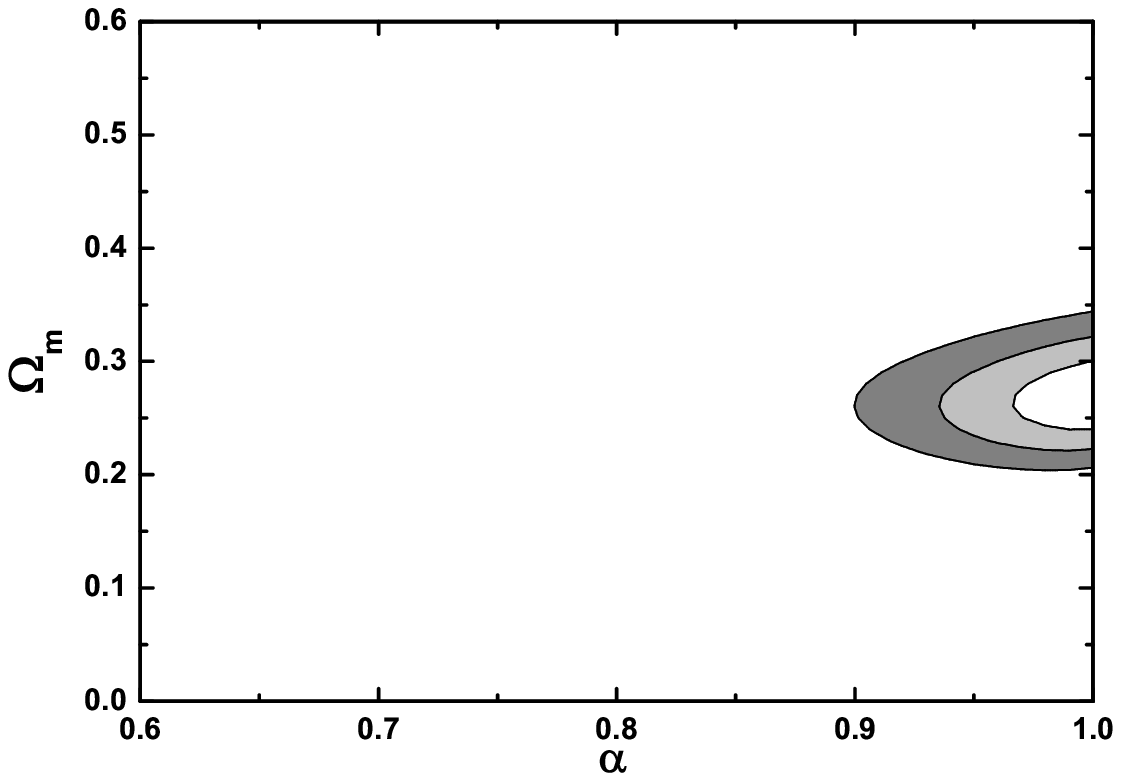,width=3.4truein,height=2.8truein}
\hskip 0.1in} \caption{Confidence contours on the space parameter
($\Omega_{\rm M}, \alpha$) from a joint analysis involving SNe Ia and the SDSS baryon acoustic oscillations. As in
Figure 2, panel (a) display the results for the \emph{gold} sample while Panel (b) for the SNLS collaboration. The important aspect here is that the parameter $\alpha$ of the simplified model is extremely restricted by both set of data.}
\end{figure*}

We estimated the best fit to the set of parameters $\mathbf{p}$ by
using a $\chi^{2}$ statistics
\begin{equation}
\chi^{2} = \sum_{i=1}^{N}{\frac{\left[\mu_p^{i}(z|\mathbf{p}) -
\mu_o^{i}(z|\mathbf{p})\right]^{2}}{\sigma_i^{2}}},
\end{equation}
with the parameters $\Omega_{\rm{M}}$ and $\alpha$ spanning the
interval [0,1] in steps of 0.01. In the above expression, $N =
157$ and $115$ for \emph{gold} and SNLS samples, respectively,
$\mu_p^{i}(z|\mathbf{p})$ is given by Eq. (\ref{dm}),
$\mu_o^{i}(z|\mathbf{p})$ is the extinction corrected distance
modulus for a given SNe Ia at $z_i$, and $\sigma_i$ is the
uncertainty in the individual distance moduli. In our analysis,
$H_0$ is considered a \emph{nuisance} parameter so that we
marginalize over it.

In Figures (1b) and (1c) we show the results of our statistical
analysis. Contours of constant likelihood (99.73$\%$, 95.4$\%$ and
68.3$\%$) are shown in the parametric space
$\alpha-\Omega_{\rm{M}}$. Panel (1b) displays the results for the
HZS \emph{gold} sample. Compared to Fig. 4 of Ref. \cite{alcaniz},
the parameter $\alpha$ is now considerably more restricted than in
the standard C-gas approach.  In particular, note that for any
value of the matter density parameter, models with $\alpha
\lesssim 0.63$ are ruled out at 99.73\% level. The best-fit model
for this analysis occurs for $\Omega_{\rm{m}} = 0.0$ and $\alpha =
0.79$ with $\chi^{2}_{\rm{min}}/\nu = 1.13$ ($\nu \equiv$ degrees
of freedom). At 95.4\% c.l. we also find $\Omega_{\rm M}
\leq 0.36$ and $0.71 \leq \alpha \leq 1.0$. Panel (1c) shows a
similar analysis for the SNLS data. The best-fit parameters in
this case are $\Omega_{\rm{M}} \simeq 0.2$ and $\alpha = 0.96$
with $\chi^{2}_{\rm{min}}/\nu = 1.0$. Note that, when compared
with recent dynamical estimates of $\Omega_{\rm M}$, this latter
value for the matter density parameter seems to be more realistic
than the one provided by the \emph{gold} sample analysis. The SNLS
sample also imply $\Omega_{\rm M} \leq 0.34$ and $0.75
\leq \alpha \leq 1.0$ at 95.4\% (c.l.).  

\subsection{SNe Ia + LSS analysis}

The recent detection of a peak in the  large scale correlation
function at 100$h^{-1}$ Mpc separation \cite{bao} provide not only
a  remarkable confirmation of the big bang cosmology but also a
kind of ``ruler" with which cosmological scenarios can be tested.
The peak detected (from a  sample of 46748 luminous red galaxies
selected from the SDSS Main Sample) is predicted to arise
precisely at the measured scale and is basically due to baryon
acoustic oscillations (BAO) in the primordial baryon-photon plasma
prior to recombination. Here, this measurement is characterized by
\begin{eqnarray}
 {\cal{A}} \equiv {\Omega_{\rm{M}}^{1/2} \over
 {{\cal{H}}(z_{\rm{*}})}^{1/3}}\left[\frac{1}{z_{\rm{*}}}
 \Gamma(z_*)\right]^{2/3}  = 0.469 \pm 0.017, 
\end{eqnarray}
where $z_{\rm{*}} = 0.35$ is the redshift at which the acoustic
scale has been measured, and $\Gamma(z_*)$ is the dimensionless
comoving distance to $z_*$.

The dotted lines in Figs. (1b) and (1c) represent the constraints
from SDSS BAO measurements on the parameter space $\Omega_{\rm{M}}
- \alpha$. Note that they are approximately orthogonal to those
arising from SNe Ia data, which indicates that possible
degeneracies in the $\Omega_{\rm{M}} - \alpha$ plane may be broken
from a joint analysis involving these observational data sets.
This is exactly what we show in Panels (2a) and (2b) for the
BAO+\emph{gold} and BAO+SNLS samples, respectively. Note that the
available parametric plane in both cases is considerably reduced
relative to the former analyses (Figs. 1b and 1c). Note also that, although compatible with the data, the region $\alpha
> 1$ (forbidden from thermodynamic
stability and causality considerations) should be disregarded from the analysis since these arguments lead to the physical bound $0 \leq \alpha \leq 1$. For
the BAO+\emph{gold} sample we find $\Omega_{\rm{M}}=
0.28^{+0.043}_{-0.048}$ and $\alpha \geq 0.916$ (with the best fit
$\alpha=0.98$) at 95.4\% (c.l.) while for the BAO+SNLS sample the
best-fit model happens at $\Omega_{\rm{M}} = 0.27$ and $\alpha =
1.0$. This latter best-fit scenario corresponds to an accelerating
universe with $q_0 \simeq -0.5$, a total age of the Universe of
$t_o \simeq 10.2h^{-1}$ Gyr, and a D/A redshift transition (from
deceleration to acceleration) $z_{\rm{D/A}} \simeq 0.75$. At
95.4\% c.l., the BAO+SNLS analysis also provides $0.94 \leq \alpha
\leq 1.0$ and $\Omega_{\rm M}=0.27^{+0.048}_{-0.045}$.

\section{Discussion and Conclusions}

As widely known, there are many theoretical approaches for
describing the exotic dark energy component accelerating the
Universe. However, until the present, the available battery of
cosmological tests was not capable to decide which is the best
theoretical representation. We have argued here that one of such
candidates, the so-called Chaplygin type gas (whose equation of
state depends on two parameters $A_s$ and $\alpha$), may have a
very simplified description. We postulate that $A_s$ is a function
of $\alpha$ and for simplicity we have taken $A_s = \alpha$. Thus, similarly to the concordance model ($\Lambda$CDM), the
resulting flat cosmology is completely described only by a pair of
parameters ($\alpha$, $\Omega_{\rm M}$).  This SC-gas cosmology
mimics the dynamics of the X-matter models  with an extra bonus,
namely: the fluid stability and other thermodynamic features are
guaranteed from the very beginning.

By considering this particular parameterization we have investigated constraints on the $\alpha$ parameter from the most recent SNe Ia (\emph{gold} and SNLS samples) and LSS data. We have found that the limits arising from this particular combination of the data are much more restrictive on this simplified approach than on the generalized C-gas version. In particular, for the the BAO+SNLS combination we found $0.94 \leq \alpha
\leq 1.0$ and $\Omega_{\rm M}=0.27^{+0.048}_{-0.045}$ (at 95\% c.l.), which is in agreement with recent estimates of the clustered matter. Naturally, it should be interesting to investigate whether current CMB data and other independent observations can or cannot discard the simplified scenario proposed here.

\begin{acknowledgments}
This work is partially supported by the Conselho Nacional de
Desenvolvimento Cient\'{\i}fico e Tecnol\'{o}gico (CNPq - Brazil).
JASL and JVC thanks FAPESP (Funda\c{c}\~{a}o de Amparo \`a
Pesquisa do Estado de S\~ao Paulo - Brazil). JSA is also supported
by Funda\c{c}\~ao de Amparo \`a Pesquisa do Estado do Rio de
Janeiro (FAPERJ) No. E-26/171.251/2004.

\end{acknowledgments}

\end{document}